\documentclass[aps,pre,reprint,amsmath,groupedaddress]{revtex4-1}
\usepackage{txfonts}
\usepackage{bm}
\usepackage{graphicx}
\usepackage{color,ulem}

\def\Vec#1{\mbox{\boldmath $#1$}}

\begin{document}

\title{Compressive response and helix formation of a semiflexible polymer
confined in a nanochannel}
\author{Yumino Hayase$^1$}
\email{yumino@hiroshima-u.ac.jp}
\author{Takahiro Sakaue$^{2,3}$, and Hiizu Nakanishi$^2$}
\affiliation{
$^1$Department of Mathematical and Live Sciences, Hiroshima University, Hiroshima 739-8526, Japan \\
$^2$Department of Physics, Kyushu University, Fukuoka 819-0395, Japan\\
$^3$ JST, PRESTO, 4-1-8 Honcho Kawaguchi, Saitama 332-0012, Japan} 

\begin{abstract}
Configurations of a single semiflexible polymer is studied when it
is pushed into a nanochannel in the case
 where the polymer persistence length $l_p$ is much longer than
 the channel diameter $D$, i.e.  $l_p/D \gg 1$.
Using numerical simulations, we show that the polymer undergoes a
sequence of recurring structural transitions upon
longitudinal compression, i.e. random deflection along the channel,
helix going around the channel wall, double-fold random deflection,
double-fold helix, etc.  We find that the helix transition can be
understood as buckling of deflection segments, and the initial helix
formation takes place at very small compression with no appreciable
weak compression regime of the random deflection polymer.  
\end{abstract}


\maketitle

\section{introduction}
The behavior of semiflexible polymers in confined spaces is
interesting for various reasons. Consider a polymer with the persistence
length $l_p$ and the contour length $L$, which is confined in the space
with the characteristic size $D$. The degree of confinement can be
quantified by the ratio $D/R$, where $R$ is the natural size of the
chain in bulk, i.e. $R \simeq l_p (L/l_p)^{1/2}$ for an ideal
chain. For polymers which are flexible down to the scale of molecular
thickness $a$, i.e. $l_p/a \sim 1$, it is indeed possible to construct
a scaling theory on the behavior of confined polymer based solely on the
ratio $D/R$~\cite{deGennes_book,Grosberg_book,Sakaue_Raphael2006}. In
contrast, the large persistence length $l_p/a \gg 1$ in semiflexible
polymers introduces an additional measure $D/l_p$ for the degree of
confinement, which leads to a rich variety of scenarios both in
nanochannel/slit~\cite{Chen06, Odijk2008,Reisner12,Cifra12,  Dorfman2011,Huang14, Werner15} and in closed
cavity~\cite{Sakaue2007, Kloz15}.  Examples of current hot topics include the
elongation of DNA in nanoscale channels~\cite{Reisner12} and the
constrained dynamics of actin filaments and microtubles in cytoskeletal
network~\cite{Koester05, Gerdel2008}. The former is becoming
an indispensable technique in single molecule genomics, and the latter
largely dictates the rheological behavior of cells.
In such fields of single molecule biophysics, a rapid progress in
molecular visualization and manipulation techniques to smaller and
smaller length scale is continuously stimulating the study on statistical
mechanical description of confined semiflexible polymers.

In this report, we consider the situation, in which a
semiflexible polymer confined in a nanochannel is compressed by a
sliding piston as shown in Fig. \ref{fig:snap}.  If the polymer is
flexible at the scale of confinement, i.e. $l_p/D \ll 1$, the polymer is
compressed randomly, developing a dense pile of blobs analogous to those in a semidilute
solution~\cite{Sakaue_Raphael2006, Jun_2008}.  Our focus here is on the
opposite limit $l_p/D \gg 1 $, where a random folding state is
energetically disfavored. Although the statistics of
a semiflexible polymer in such a narrow channel  is
now rather well understood as Odijk regime~\cite{Odijk83}, the
response to the compressive force has not been addressed yet. Using
Langevin dynamic simulation, we show that in such a case the polymer
transforms from a random deflection configuration into an ordered
helix structure upon compression.  Further compression leads to
destabilization of the helix into a double-fold random deflection,
then the second buckling takes place to form the double-fold helix. We expect that such a sequence of
recurring structural transitions is
within reach using {\it nano-piston} experiment set-up~\cite{Khorshid14,
Khorshid2016, Reccius05, Pelletier12}, and its investigation offers interesting challenges to
explore rich confinement scenarios realized in the halfway between
nanochannel and closed cavity geometries, where the effective spacial
dimensionality changes from one to zero.
 
\section{model}
In numerical simulations, we employ a coarse-grained model of semiflexible
polymer, which is made of $N$ successive beads with diameter
$\sigma$. All beads interact through a shifted purely repulsive
Lennard-Jones potential $U^{LJ}$,
\begin{equation}\label{U_LJ}
U^{LJ} (r) = \begin{cases}\displaystyle
4 \epsilon \left[ \left\{ \frac{\sigma}{r} \right\} ^{12} - 
\left\{ \frac{\sigma}{r} \right\} ^6 +
	      \frac{1}{4} \right]  & r  < 2^{{1}/{6}} \sigma \\
0 & r  > 2^{{1}/{6}} \sigma
\end{cases},
\end{equation}
where $r$ is the inter-bead distance, and $\epsilon$ sets the energy scale.  
The linear connectivity of the chain is ensured by the
Finitely-Extensible-Nonlinear-Elastic potential $U^{s}$ 
between neighboring beads \cite{Grest89},
\begin{eqnarray}
U^{s} (r)
= -\frac{1}{2} k_s r_0^2 
\ln{\left[ 1- \bigg(\frac{r}{r_0}\bigg)^2\right]} .
\end{eqnarray}
Finally, the stiffness of the chain is controlled via the bending potential
\begin{equation}
U^{b} (\theta) =\frac{\kappa}{\sigma}\,(1-\cos\theta) , 
\end{equation}
with $\theta$ being the angle between two consecutive bonds along the
chain. The chain therefore approximates the worm-like chain with the
persistence length $l_p = \kappa/k_BT$, where $k_BT$ is the thermal
energy.

The chain is confined in the cylindrical channel with two end caps,
whose size is characterized by its cross sectional diameter $D$ and the
axial length $X$.  To compress the chain, the cap at
$x=X(t)$ moves with a constant velocity $c$ in the $-x$-direction
with the cap at $x=0$ being fixed, where we set the channel axis
to be the $x$-axis.  The surface of such a confining geometry is
implemented via the potential $U^c$, 
for which we also use $U^{LJ}$ of Eq. (1). 
Note that the distance $r$ in this case is form the cylinder wall, whose diameter
is set to be $D+2\sigma$ so that the effective diameter for the beads should be $D$.

The position ${\Vec r}_i = (x_i, y_i, z_i)$ of the $i$-th bead
evolves with time according to Langevin equation
\begin{equation}
m  \frac{d^2 {\Vec r}_i}{dt^2} 
= - \nabla_i U - \Gamma \frac{d {\Vec r}_i}{dt} + {\Vec W}_i(t) , 
\label{LangevinEq}
\end{equation}
where $m$ and $\Gamma$ are the bead mass and the friction coefficient,
respectively, and $U$ is the sum of all the potential 
$U=\sum_{i.j}U^{LJ}+\sum_i \left(U^s+U^b+U^c\right)$. 
The random force ${\Vec W}_i(t)$ is Gaussian white noise with
zero mean and the covariance
$\langle W_{i \alpha}(t) {W}_{j \beta}(t') \rangle 
=  \delta_{\alpha \beta} \delta_{i j} \delta(t-t') 2 k_B T  \Gamma $,
where $\alpha$ and $\beta$ represent $x$, $y$, or $z$.

We take $\sigma$, $\epsilon$, and $m$ as basic units, which set the
time unit $\tau\equiv \sigma\sqrt{m/\epsilon}$.
The parameters we used are $r_0=1.5 \sigma$, $k_s=30 \epsilon/\sigma
^2$, $k_B T=0.1 \epsilon$ and $\Gamma =m/\tau$.  The
polymer with $N=512$ beads is placed in the cylindrical channel with the
 diameter $D=5 \sigma$. The potentials $U^{LJ}$ and $U^s$
keep the bond length $b$ nearly constant with the average $\langle b
\rangle \simeq 0.96 \sigma$, which leads to the polymer contour length
$L_0 = (N-1) \langle b \rangle \simeq 491 \sigma$. Unless otherwise
stated, we set $\kappa = 50 \epsilon \sigma$, which corresponds to $l_p = 500
\sigma$, namely, $l_p\approx L_0$.
We numerically integrate Eq.~(\ref{LangevinEq}) with the time step
$\delta t = 0.008 \tau$.

\section{numerical results}
\begin{figure}
\includegraphics[width=6cm]{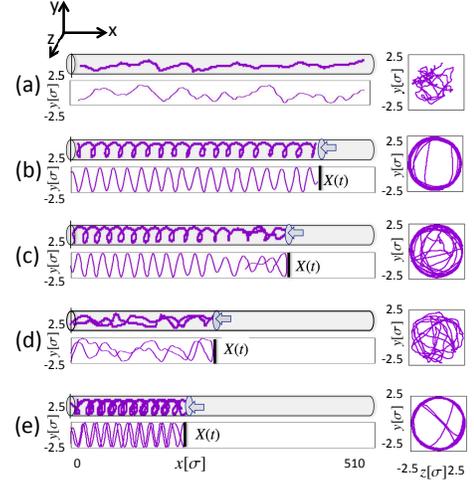} \caption{Typical snapshots of a
(un)compressed polymer chain in the cylinder of the length
$X(t)=510\sigma$ (a), $414\sigma$ (b), $362.32\sigma$ (c), $238\sigma$
(d), and $190\sigma$ (e). In each case, the three-dimensional
configuration (top), and its projection onto the $x$-$y$ plane
(bottom) and the $y$-$z$ plane (right) are shown.
Because the aspect ration of the system is to large,  the scale of the $x$-coordinate is different from those of the $y$ and $z$-coordinates.  
}
\label{fig:snap}
\end{figure}

To prepare initial conditions, we first align the beads along the
cylinder axis at ${\Vec r}_i = (i \sigma, 0, 0)$,
and run simulations until the chain reaches the equilibrium state in the
channel without end caps. As shown in Fig. \ref{fig:snap} (a), the chain
is globally oriented along the channel with apparent random
deflections, whose characteristic undulation
mode is determined by the interplay among the thermal fluctuation,
the bending elasticity and the confinement effect.  The measured
extension of the chain along the $x$-axis
in this uncompressed state $\langle L_x
\rangle=\langle x_N-x_1 \rangle =487.45 \sigma$ is in quantitative
agreement with $487.1 \sigma$ obtained by the expression for the Odijk regime
\begin{eqnarray}
\langle L_x \rangle = L_0[1- \alpha_{0} (D/l_p)^{2/3}]
\label{L_x_eq}
\end{eqnarray}
with a geometry-dependent numerical constant $\alpha_{0} \simeq
0.17$ for a channel with a circular cross
section~\cite{Yang07}.  After
the chain reached the equilibrium state, we start to compress the
chain by sliding the end wall at $X (t)$ slowly with the
velocity $c=10^{-4} (\epsilon/m)^{1/2}$. Smaller speeds have also
been tested to confirm that the transformation scenario described below
does not change ~\footnote{The time scale for a  nanometer size particle in water diffusing over its own size is on the oder of nanosecond. Assuming the monomer size to be $1 \sim 10$ nm, its diffusion time $\Gamma \sigma^2 /k_{B}T$ is roughly estimated as $10$ ns. This leads to the order of magnitude for the velocity scale $\sqrt{\epsilon/m} \sim 1$ m/s, hence, the speed $c \sim 10^2 \ {\rm \mu m /s}$ of nano-piston in simulation.
}.
Upon being compressed, the chain changes its configuration from a
random deflection to a helix; a snapshot at
$X(t)=414 \sigma$ is shown in Fig. \ref{fig:snap}(b).  Further
compression causes one of the chain ends to turn around as in
Fig. \ref{fig:snap} (c). This entails rapid decrease in the pressure at
both of the walls, as the helix relaxes. The kink, i.e.  turning point,
moves to the middle of the chain as the piston moves, and the chain becomes
doubly folded as in Fig.  \ref{fig:snap} (d). At a higher
compression, the double-fold chain transforms to
a double-fold helix as shown in Fig.\ref{fig:snap} (e).  Eventually, the
double helix turns around once more and relaxes as the single helix does.

\begin{figure}
\includegraphics[width=4cm, angle=-90]{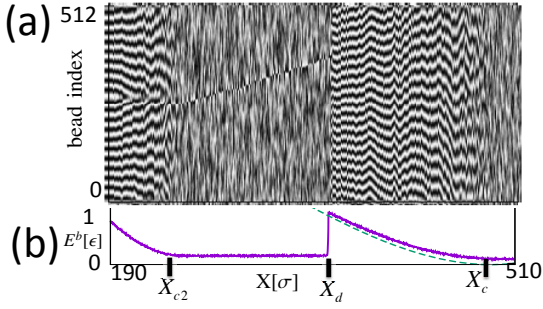} \caption{(a) Kymographic
representation of the chain configuration characterized by the sequence
of $y$ coordinate of beads. Horizontal and vertical axes specify the
channel length $X$ and the bead index $i$, respectively, 
The $y$ -coordinate is shown in the gray scale.
(b) Average bending energy per bending angle
as a function of $X$. The dashed line represents the analytic expression for noise-free structure given in Eq.~(\ref{E_bend_0}). }
\label{fig:st}
\end{figure}

Figure \ref{fig:st}(a) shows the change in polymer configuration upon
compression; the $y$-coordinate of bead positions $y_i(t)$ are plotted
in the gray scale for the bead number $i$ on the vertical axis and
the distance between the walls $X(t)$ on the horizontal axis.  Note that
the initial randomly deflected configuration is represented along the
vertical line at the right end of the plot at $X=510 \sigma$, and the
system moves towards left as being compressed. The randomly blurred
patterns in the plot correspond to the random deflection.
Around $X (t)=X_c \simeq 488 \sigma$ the helix structure is formed as
can be seen from the zebra pattern. 
It should be noted that $X_c \simeq \langle L_x \rangle$, i.e., the helix transition takes place under the minute influence of the piston. We will discuss this later.

  As $X(t)$ decreases further, the pitch of the helix becomes shorter.  At $X(t)=X_d
\simeq 365 \sigma$ in Fig. \ref{fig:st}(a), the right end of the chain turns around and the helix
structure vanishes as the chain relaxes.  The sharp line appearing in the
blur region for $X_{c2}<X<X_d$ shows the bead number where the chain
turns around.
One can see that the kink appears at $X=X_d$ near the end of
the chain, then it moves towards the middle of the
chain.
At $X(t)=X_{c2} \simeq 245 \sigma$, a double-fold helix is formed, whose
pitch decreases upon further compression.

The bending energy $E^b= \frac{1}{N-2}\sum_{i=2}^{N-1}
U^b_i$ is shown in Fig. \ref{fig:st}(b). As the chain is compressed,
$E^b$ increases, but decreases discontinuously
at $X(t)=X_d$ when one of the chain ends turns around,
because the helix relaxes into the double-fold random
deflection state. 
 In the region $X_{c2} < X (t) <X_d$ the bending energy $E^b$ stays low and
starts to increase again when the double-fold helix is formed.

\begin{figure}
\includegraphics[width=5cm,angle=-90]{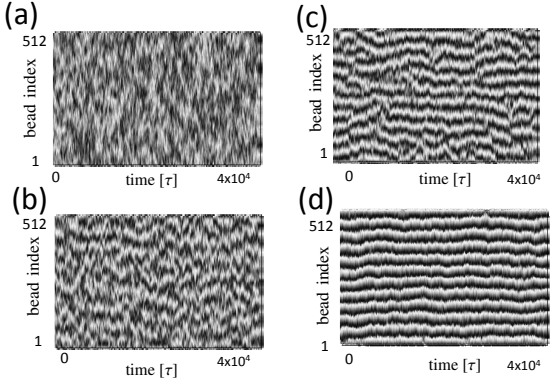}
\caption{Kymographic representation of the spatio-temporal
 configurations of the chain at 
$X =498\sigma$ (a),  $483\sigma$ (b), $475\sigma$ (c), and $467\sigma$ (d). 
The $y$ co-ordinate is shown in the gray scale.
}
\label{fig:helix}
\end{figure}

Fig. \ref{fig:helix} shows the spatio-temporal structural patterns
with the channel length $X$ being fixed at some values near $X_c$.
Unlike Fig.~\ref{fig:st}(a), the horizontal axis in
Fig. \ref{fig:helix} is time, so that the time course of the
fluctuating structures near the onset of helix formation are clearly
visible.  In Fig.~\ref{fig:helix}(a) at $X = 498 \sigma > \langle L_x \rangle=487.45
\sigma$, one only sees a disordered pattern,  
which corresponds to the configuration of an uncompressed confined polymer spreading
along the cylinder axis (Fig.\ref{fig:snap}(a)).  In
Fig. \ref{fig:helix}(b) at $X=483\sigma < X_c \simeq 488 \sigma$, a
highly fluctuating  but characteristic zebra structure is
visible, indicating that the chain has helical turns, which are created
and annihilated temporally.  With further compression, the
fluctuating helical structures become more stable
 at $X=475 \sigma$ in Fig. \ref{fig:helix}(c), and the helix structure
establishes through the whole chain at $X=467 \sigma$ in
Fig. \ref{fig:helix}(d).

\section{discussion}
The characteristic force of the helix formation could be addressed through the derivative of the bending energy, i.e., $f_c = L_0 |d E^b(X)/d X|_{X=X_c}$ at the transition point. To evaluate it,  we consider the helix configuration without thermal noise, which is  represented as 
\begin{eqnarray}
{\Vec r}_{\natural}(s) = \left( \frac{ps}{\pi D} \sin{\phi}, \frac{D}{2} \cos{\theta (s)}, \frac{D}{2} \sin{\theta (s)}  \right)
\label{r_s_0}
\end{eqnarray}
with the arc length $s \in (0, L_0)$, where the subscript ($\natural$) indicates the quantity without noise.
Here the angle $\phi$ is related to the helix pitch $p$ through $\tan{\phi} = \pi D/p$, and $\theta(s) = 2 s \sin{\phi}/ D$ is the angle in the cross-sectional circle.
Noting the axial length of the helix $X = L_0 \cos{\phi}$, the bending energy of the structure is calculated as
\begin{equation}
E^{b}_{\natural}(X)=\frac{\kappa}{2L_0} \int^{L_0}_0 ds \left(\frac{d^2 \bm r_{\natural}(s)}{ds^2} \right)^2 = \frac{2 \kappa }{D^2}\left[ 1- \left(\frac{X}{L_0} \right)^2\right]^2
\label{E_bend_0}
\end{equation}
From this and using the relation $X_c \simeq \langle L_x \rangle$ given in Eq.~(\ref{L_x_eq}), we find
\begin{equation}
f_{c \natural} 
\approx 16 \alpha_0 {\kappa\over\lambda^2},
\label{f_c}
\end{equation}
where $\lambda = l_p^{1/3}D^{2/3}$ is the deflection length~\cite{Odijk83}. Apart from a numerical constant $16 \alpha_0 \simeq 2.7$, this expression coincides with the buckling force of the filament with the bending rigidity $\kappa$ and the length $\lambda$~\cite{Landau_Elasticity}. Our analysis thus identifies the helix formation as the buckling of the deflection segment.

In reality, the helix is substantially disturbed by the thermal noise, in particular, close to the transition point. We thus analyze the spatial correlation near the onset of the helix formation
$X \simeq X_c$ by the orientational correlation function
\begin{equation}
C_b (|i-j|) =\langle \Vec{t}^{\perp}_i (t) \cdot \Vec{t}^{\perp}_j (t)\rangle, 
\end{equation}
where $\Vec{t}_i^\perp$ is the transverse component of the tangent
vector $ \Vec{t}_i (t) \equiv
(\Vec{r}_{i+1}-\Vec{r}_i)/|\Vec{r}_{i+1}-\Vec{r}_i| \equiv \Vec{t}^x_i
(t)+\Vec{t}^{\perp}_i(t) $.  For the uncompressed state, one can
evaluate the correlation function analytically by the effective
free energy~\cite{Koester05, Frey2007,Wang2007}
\begin{equation}
F=\frac{\kappa}{2} \int^{L_0}_0 ds \left(\frac{d^2 \bm r(s)}{ds^2} \right)^2+\frac{k}{2} \int^{L_0}_0 ds (\bm r ^{\perp}(s))^2,
\end{equation}
where the fluctuating chain configuration is represented as ${\Vec r}(s)$ with the arc length $s \in (0, L_0)$ so that the tangent vector is given by ${\Vec t}(s) = d{\Vec r}(s)/ds$.
The confinement effect is modeled
by the harmonic potential with the spring constant per unit length $k$. The
excluded-volume effect can be safely neglected in Odijk regime
$D/l_p \ll 1$, where small fluctuations around the straight
configuration can be analyzed by the Gaussian approximation, leading to
\begin{eqnarray}
\langle {\Vec t}^{\perp}(s) \cdot {\Vec t}^{\perp}(0) \rangle = \frac{2}{L_0}\frac{k_BT}{\kappa} \sum_{q_n}\frac{q_n^2}{q_n^4+ k/\kappa} e^{-iq_ns} 
\label{bond_c_1}
\end{eqnarray}
with the wave number $q_n = 2 \pi n/L_0$.  Approximating the summation
in Eq.~(\ref{bond_c_1}) by the integral, we obtain 
\begin{equation}
\langle {\Vec t}^{\perp} (s) \cdot {\Vec t}^{\perp} (0) \rangle =\frac{\lambda}{\sqrt{2} l_p}  \cos \left( \frac{s}{\lambda}+\frac{\pi}{4} \right) \exp \left( -\frac{s}{\lambda} \right), \label{bond_c}
\end{equation}
where the deflection length $\lambda = l_p^{1/3}D^{2/3} = (4 \kappa/k)^{1/4}$  naturally appears from the requirement that the chain is confined in a cylinder with diameter $D$, i.e., $\langle |{\Vec r}^{\perp}(s)|^2 \rangle = (D/2)^2$.

\begin{figure}
\centerline{ \includegraphics[width=4cm,angle=-90]{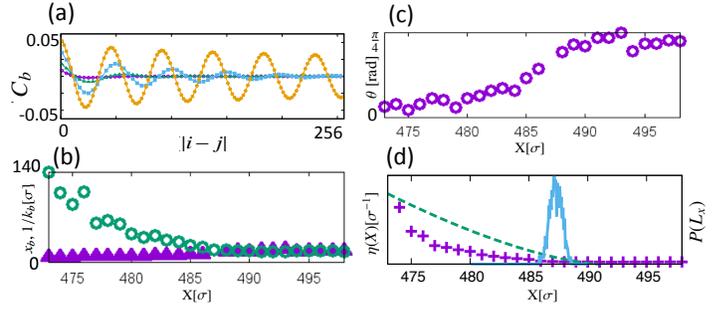} } \caption{
(a) The bond correlation function
 $C_b$ is plotted in the vicinity of $X_c$ 
for $X=500\sigma$ ($\circ$), 
$485\sigma$  ($\blacktriangle$), 
$477\sigma$  ($\blacksquare$), and
$469\sigma$ ($\bullet$) with $l_p=500$. 
(b)Characteristic decay lengths $x_b$ ($\circ$) and
inverse wave number $1/k_b$ ($\blacktriangle$) of the bond correlation
function as a function of $X$. 
 (c) Phase sift $\theta$ in
the bond correlation function as a function of $X$.  
(d) Helical order
parameter $\eta$ as a function of $X$: the numerical results (+) and the analytical expression (\ref{eq:oder_ana}) without thermal noise (broken line).
 The probability distribution $P$ of the projected chain length $L_x$ without
compression is overlaid with the common $x$ axis (blue line).}  
\label{fig:new}
\end{figure}

Fig. \ref{fig:new}(a) shows the correlation function $C_b(|i-j|)$ obtained by simulations
for various channel lengths $X$.  The functions
are well fitted to the formula 
\[
 C_b(|i-j|) = A_b \cos(k_b |i-j| +\theta) \exp (-|i-j|/x_b).
\]
The fitting parameters are shown in
Fig. \ref{fig:new}(b) and (c) in the vicinity of $X=X_c$.  For large
 $X$, $x_b$ and $k_b$ are almost the same and nearly
constant $x_b \simeq 1/k_b \simeq 20\sigma$ in agreement with the
theoretical estimate by Eq.~(\ref{bond_c}), which gives $\lambda \simeq
23 \sigma$ for $D=5 \sigma$ and $l_p = 500 \sigma$. The deviation from
this at $X < X_c$ should be the compression effect.  Upon decreasing $X$
below $X_c$, the pitch of the helix $1/k_b$ decreases while the
correlation length $x_b$ increases rather rapidly, which means that the
helix structure prevails.  The phase shift $\theta$ in
Fig. \ref{fig:new}(d) takes $\theta = \pi/4$ in the large $X$ limit,
 which is consistent with Eq. ~(\ref{bond_c}). Under compression, $\theta$
decreases toward $\theta=0$.

We introduce the helical oder parameter $\eta$ defined by
\begin{equation}
\eta = \frac{1}{\sigma}  \langle   {\Vec t}^{\perp}_{i} \times {\Vec t}^{\perp}_{i+1}  \rangle 
\end{equation}
which becomes large for the helical structure. 
If there is no thermal fluctuation, one can again use Eq.~(\ref{r_s_0}) to calculate the helical order parameter
\begin{equation}
\eta_{\natural}(X) = \frac{2}{D} \left(\frac{L^2-X^2}{L^2}\right)^\frac{3}{2} \approx \frac{2}{D} \left(2 \frac{L-X}{L} \right)^\frac{3}{2}.
\label{eq:oder_ana}
\end{equation}
In Fig. \ref{fig:new}(d), we shows the helical order
parameter  obtained by the numerical simulation along with the analytical expression (\ref{eq:oder_ana}) for the case without fluctuation. 
The oder parameter $\eta$ of the numerical results starts to grow at $X \simeq X_c$. 
The chain seems to form the helix right at the onset of compression
with no appreciable weak compression regime without any structural change. 
Upon further compression, the oder parameter reaches to the value for the no thermal fluctuation case.
To resolve the onset point more carefully, 
we overlay in Fig. \ref{fig:new}(d) the distribution
of the chain extension $L_x$ without the compression, from which the
width $\sqrt{(\delta L_x)^2} \simeq 0.516 \sigma$ is extracted. A
careful inspection of Figs. \ref{fig:new}(a)-(d) shows  $X_c \simeq
\langle L_x \rangle $, thus, the chain feels the
compression force and starts to form a helix when the channel
length reaches the average chain extension in the channel.

Such a feature of the compressive response, i.e., sudden buckling into
helix without a notable sign of ordinary linear response regime, could
be understood in the following way.
The fluctuation of $L_x$ has been estimated as
\begin{eqnarray}
 \langle (\delta L_x )^2\rangle
 = \beta_0 \frac{D^2}{l_p} L_0
 = \beta_0 \frac{\lambda^3}{l_p^2}L_0
 \label{deltaL_x_eq}
 \end{eqnarray}
with $\beta_0 \simeq 0.00754$ for
the circular channel~\cite{Gompper2010}.  The standard deviation
by this expression
$\sqrt{\langle (\delta L_x )^2\rangle}=0.43 \sigma$ is in agreement with
$0.516 \sigma$ by the numerical simulation 
\footnote{The standard deviation by the numerical simulation $0.516 \sigma$ is slightly larger than the value estimated in Eq. (\ref{deltaL_x_eq}) ,i.e. $0.43 \sigma$. It is because Burkhart et al. estimated  $\beta_0$ by using  the hard wall potential in \cite{Gompper2010}. They notice that $\beta_0$ becomes larger for the wall with soft potential as in the present study. We also expect that the details of the simulation model, such as the type of potentials, may affect the precise location of the transition point, but the deformation scenario does not change.For instance, the transition points $X=X_{c,d,2c}$ may depend on the manner of the chain discretization. 
}.
If we estimate the spring constant as
\begin{eqnarray}
k_{\parallel}
 = \frac{k_BT}{\langle (\delta L_x )^2\rangle},
\end{eqnarray}
the typical force $f^*$ in the linear response regime should be
\begin{eqnarray}
f^* \simeq k_{\parallel} \sqrt{\langle (\delta L_x )^2\rangle}
 = \frac{k_BT}{\sqrt{\langle (\delta L_x )^2\rangle}}.
\end{eqnarray}
On the other hand, as we already discussed, the critical force $f_c$ to induce the helix transition is identified as the buckling force
of the deflection segment as given by Eq.~(\ref{f_c}). 

In order for the weak compression regime to exist, we would expect
$f^*<f_c$, but this condition leads to
\begin{equation}
 L_0 \ > \  L_0^{\rm th} \equiv {\lambda\over (16\alpha_0)^2\beta_0}
\approx {\lambda\over 0.055}.
\end{equation}


In other words, the polymer with $L_0$ shorter than $L_0^{\rm th}$ will
transform into helix by the compression weaker than that by the thermal fluctuation.
Because $\beta_0$ is very small, $L_0^{\rm th}$ could be
quite large; for our chain with $\lambda \simeq 23
\sigma$, $L_0^{th} \simeq 420 \sigma$, already comparable to the chain length
$L_0 \simeq 491 \sigma$ used in our simulation~\footnote{This concrete number is based on our estimation of $f_{c\natural}$ (Eq.~(\ref{f_c})) without thermal fluctuation effect. One may expect the fluctuation may facilitate the buckling, i.e., $f_c < f_{c \natural}$, which yields even larger value for $L_0^{\rm th}$.} . In principle, the compressive mode of a long
chain is very soft (cf. $L_0^{-1}$ factor in the spring
constant). However, the smallness of $\beta_0$ makes the stiffness
relatively high for practical chains with moderate length. This
stiffening facilitates rigid response to the compressive force, in
which the deflection segment buckles into the helix.
We have checked that the scenario described above is robust against the
change in the persistence length. For chains with $l_p /\sigma = 250, \
100$, the helix formation at $X_c$ and its collapse into the double-fold
at $X_d$ are clearly seen, albeit with the shift of the critical values
$X_c$ and $X_d$. For chains with $l_p/\sigma=50$, the helical structure
in the snapshot is not easily recognizable by the naked eye; the chain is
transformed into the double-fold before the clear helix is
formed. But the analysis of the spatial correlation $C_b$
and the order parameter $\eta$ indicates the departure from the
disordered structure in a certain range of $X$, hinting the local
and transient helix. 
Overall, with the decrease in $l_p/D$, the interval for the helix state $X_c -X_d$ decreases, and it is expected that such an interval would eventually disappear for sufficiently small $l_p/D$. Our simulation indicates the threshold $(l_p/D)_{th} \sim 10$ i.e. $(\lambda/D)_{th}=(l_p/D)^{1/3} \sim 2 $.
We do not understand yet the mechanism of helix instability at
$X=X_d$.

If we move the end cap backward after the double-fold state
($X_{c2} < X < X_d$) is formed, the helix never forms again,
but the uncompressed state is reached directly.  This
hysteresis is easy to understand by comparing the bending energies
between the helix and the double-fold states under compression. While
the bending energy of the helix increases under
compression, that of the double-fold random deflection
state is constant as long as a turing point exists.  In the case
of $l_p/D \gg (l_p/D)_{th}$, the compression of the initially unperturbed chain
always leads to the formation of helical structure because the bending
energy of nascent helix is lower than that of the double-fold
random deflection state. The helix, once formed, is very stable,
because turning around a chain end gets more unlikely for larger $l_p/D$ due to higher energetic penalty of bending.

In order to examine experimental relevance of the phenomenon, let us
discuss some of the system parameters.  Most of our simulations are
performed using a polymer with its persistent length $l_p = 500 \sigma$
pushed into the cylinder of the diameter $D=5 \sigma$, but we also performed
some simulations with $l_P=100 \sigma$ and $50 \sigma$ and observed
the helix formation.
If we use a DNA of $l_p \sim 50$ nm, a nanochannel diameter of
$D=5\sigma$ corresponds to $D= 0.5 \sim 5$ nm. In the case of stiffer
polymer such as actin filament, for which
$l_P\sim 16 \ {\rm \mu}$ m, the cylinder diameter
can be as large as sub-micron range, which should be realistic
experimentally.
The critical force of compression $f_c$ given by Eq.~(\ref{f_c}) for
the helix formation is of the order of pN for the actin filament
compressed in the cylinder of the diameter $D \sim 100$ nm, where the
deflection length $\lambda \sim 500$ nm.

The experimental set up for nano-piston has been proposed in
ref.[17,18], where the nano-piston is controlled by an optically trapped
bead.  Another possibility may be AFM, in which case the size of the
bead is not limited by the wave length of the laser.
We expect that the present process is experimentally accessible through
the hysteresis in the relation between the displacement of the piston
and the compression force during compression and extension.

\section{conclusion}
In summary, we have reported the formation of helical structure of
semiflexible polymer in a nanochannel subjected to the
longitudinal compressive force.  This would provide
an interesting possibility to control the structure of various
semiflexible polymer under confinement.  In the present report, we have analyzed 
the initial response, and have pointed out
the peculiarity of the compressive response of semiflexible polymer;  
the phenomenon is most naturally interpreted as
the buckling of deflection segment,
and the structural transition to the helix takes place even at the very weak
compression by the confinement length $X$ comparable to $\left<L_x\right>$.
This means the intermittent force at the cap by the thermal fluctuation
can induce the helix structure.

\begin{acknowledgments}
This work is supported by KAKENHI (No. 16H00804, ``Fluctuation and
Structure") from MEXT, Japan, and JST, PRESTO (JPMJPR16N5).
\end{acknowledgments}
\bibliographystyle{apsrev}


\end{document}